\documentclass[manuscript]{aastex}

\usepackage{natbib}
\usepackage{color}
\usepackage{graphics}
\usepackage{epstopdf}
\usepackage{hyperref}
\usepackage{subfigure}

\slugcomment{The Astrophysical Journal Letters}

\shorttitle{Flows and Waves in Braided Solar Coronal Magnetic Structures}
\shortauthors{Pant et al.}

\begin{document}


\title{Flows and Waves in Braided Solar Coronal Magnetic Structures}

\author{V.~Pant\altaffilmark{1}}
\affil{1. Indian Institute Of Astrophysics, Bangalore-560 034, India}
\email{vaibhav@iiap.res.in}
\author{A.~Datta \altaffilmark{1,2}}
\affil{2. HKBK College of Engineering,Bangalore 560\~045, India.}
\author{D.~Banerjee\altaffilmark{1, 3}}
\affil{3. Center of Excellence in Space Sciences, IISER Kolkata, India}


\begin{abstract}
We study the high frequency dynamics in the braided magnetic structure of an active region (AR 11520) moss as observed by High-Resolution Coronal Imager (Hi-C). We detect quasi periodic flows and waves in these structures. We search for high frequency dynamics while looking at power maps of the observed region. We find that shorter periodicites (30 - 60 s) are associated with small spatial scales which can be resolved by Hi-C only. We detect quasi periodic flows with wide range of velocities from 13 - 185 km s$^{-1}$ associated with braided regions. This can be interpreted as plasma outflows from reconnection sites. We also find presence of short period and large amplitude transverse oscillations associated with braided magnetic region. Such oscillations could be triggered by reconnection or such oscillation may trigger reconnection.   
\end{abstract}



\section{Introduction}
The understanding of the heating mechanism/s of the solar corona is one of the main challenges 
in solar physics. Two types of mechanisms are well accepted, namely impulsive heating by nanoflares 
\citep{parker88} and wave heating by dissipation of waves.\\
High Resolution Coronal Imager (Hi-C) \citep{kob2014} provided unprecedented details of active region moss at small spatial scales. It has diffraction limited spatial resolution of 0.3$\arcsec$ and cadence of $\sim$5.5 s. Hi-C has revealed many new features of the corona at small spatial scales \citep{peter13,morton13,Win14}. \citet{antiochos2003} reported weak intensity variation $\sim$10\% in active region moss over period of hours using Transition Region and Coronal Explorer (TRACE). They rule out the possibility of impulsive heating and also conjectured that high-frequency heating could be the source of observed variability.  \citet{Brooks2009} confirmed the findings of \citet{antiochos2003} using Hinode, EIS observation. They did not find strong flows and short term variability in moss region and concluded that heating is quasi-steady. High temporal and spatial resolution of Hi-C gave a new insight in understanding the mechanism of coronal nanoflare heating. \citet{Win13} studied inter-moss loops using Hi-C and reported that these cool and dense loops are the result of impulsive heating of magnitude similar to that of coronal nanoflares. \citet{testa13} have reported the variability $\sim$ 15--20 s in the active moss region at the foot points of bright hot coronal loops and attributed these as the signature of impulsive nanoflare events. Hi-C observations have revealed small scale brightening in EUV of duration 25 s and of length scale 0.68 Mm \citep{reg14}. \citet{Cirtain13} have reported the first evidence of magnetic field braiding and axial twist from Hi-C observations. They have estimated the free energy available $\sim$ $10^{29}$ ergs. The release of this energy due to magnetic reconnection can heat up the loop. A non-linear force free field reconstruction of magnetic field lines corresponding to the same field of view of Hi-C reveals the braiding and twisting of magnetic fields \citep{thalmann2014}. They estimated $\sim$ 100 times more free energy than estimated by \citet{Cirtain13}. Recently \citet{tiwari2014} has observed a subflare event at this region just after the Hi-C observations from different channels of Atmosphearic Imaging Assembly (AIA) on board Solar Dynamic Observatory (SDO).\\
Transverse oscillations in corona have been reported by \citet{tom2007} using Coronal Multi channel Polarimeter (COMP) data and  \citet{mc2011} using AIA on Solar Dynamic Observatory (SDO). They have reported the typical velocity amplitude of $\sim$5 km s$^{-1}$. After the advent of Hi-C \citet{morton13,morton14} reported transverse waves in active region moss with velocity amplitude as high as 11 km s$^{-1}$ and mean periodicity of $\sim$ 50 s. They have estimated that about 15\% of wave energy is carried to transition region from chromosphere.

In this letter we focus on high frequency dynamics of active region moss and braided magnetic region as seen from Hi-C, with particular emphasis on the reconnection sites. Recurrent  reconnections and waves both can contribute significantly to the heating. We search for the presence of flows which can be attributed as a signature of reconnection within the braided magnetic region.  

\section{Observations and Data Analysis}
Hi-C took high resolution images of the Solar corona in Fe~{\sc xii}~193 \AA~  passband. The observations were performed on 11 July 2012 at 18:52:09 UT with the cadence of $\sim$ 5.5 s and pixel resolution of $\sim$ 0.103 $\arcsec$ pix$^{-1}$ for a duration of 200 s. We used level 1.5, 4K X 4K dataset. This dataset is corrected for pointing drift, spacecraft jitter and atmospheric absorption \citep{kob2014}. We align datacube using cross correlation to remove residual drifts and to achieve sub-pixel accuracy \citep{morton13, morton14}.\\
We also use simultaneous imaging data as recorded by AIA/SDO  with  EUV narrow band ($\sim$0.6$\arcsec$ pixel$^{-1}$, 12 s cadence).
Fig.~1 (a) shows AIA 193 \AA~ full disk image. Small black rectangle shows full Hi-C field of view (FOV). Figs.~1 (b) and (d) show the zoomed view of region marked with black rectangle in Fig.~1 (a) as seen in AIA and Hi-C respectively. Black rectangle in Fig.~1(d) marks the region of interest (ROI) for further analysis. Fig.~1(e) shows the zoomed view of ROI and curved white slices which are used for creating time-distance maps. Fig.~1(f) shows zoomed view of subfield of ROI as marked in Fig.~1(e).
Due to high noise in the Hi-C images we filter each image into high and low spatial frequency components. High frequency image shows significant small scale structures therefore, we filter high frequency component further. We iterate it up to three times and the resulting high frequency image contains only uncorrelated noise. Finally, we subtract high frequency image from original image. Furthermore, in order to bring out structures at different spatial scales, we use normalized multi gaussian filter \citep{morgan2014} to filtered images. We choose width of gaussian filter to be 11, 21, 41 and 81 pixels for ROI and 21, 41, 101, 201 and 1001 pixels for Hi-C full FOV. We add gaussian filtered images of different spatial scales with equal weight to obtain a single multi gaussian filtered image. Above procedure is repeated for AIA 193 \AA~ image with width of gaussian filter to be 11, 21, 41 and 81 pixels and filtered image is shown in Fig.~1 (c). Filtered images of the corresponding images of Hi-C in Fig.~1 (middle panel) are shown in the bottom panel of Fig.~1. The movie file is available online. Left panel in movie corresponds to Hi-C original intensity images while right panel corresponds to multi gaussian filtered images.

\section {Results}
First we will focus on the distribution of power as calculated from wavelet methods to identify the locations where the high frequency dynamics are present. 
\subsection {Power and Wavelet maps}
We perform wavelet analysis at each pixel location of Hi-C ROI and corresponding FOV of AIA 193 \AA~. Fig.~2 top left panel shows the power map of Hi-C ROI in 30-60 s and filtered image of Hi-C overplotted with the pixel positions marked in green where global significance level of power is greater than 95 $\%$ confidence level \citep{torrence1998} for a white noise process (see, Fig.~2 bottom left panel). We limit ourselves to 30-60 s interval because we note that for higher frequencies (15-30 s), wavelet starts picking up noise and it is difficult to distinguish between noise and true signal variation. To limit the noise we discard isolated pixel locations and set threshold as 9 pixels on the size of detected regions over global significance level. We carry out similar analysis for AIA 193 \AA~. Cadence of AIA is low (12 s) as compared to Hi-C (5.4 s) thus the data points will be less due to finite duration of time series. Therefore we choose longer time series of 10 minutes starting from 18:50:00 UT to 18:60:00 UT for AIA 193 \AA~. We find that even after choosing longer time series significant power in shorter periodicities (30-60 s) is absent in AIA FOV (Fig.~2 bottom middle panel) while significant power is present in longer periodicities (120-180 s) as shown in Fig.~2 bottom right panel. For AIA we set the threshold as two pixels on the size of detected regions over global significance level.\\
Wavelet maps of pixel locations marked as A and B in Hi-C FOV (see Fig.~2 bottom left panel) are shown in Figs.~3 (a) and (b) respectively. Fig.~3 (a) Top panel shows original Hi-C light curve without any smoothing at position marked as A with error bars. Errors in data points are calculated as $\sqrt{0.23F + 588.4}$ \citep{morton13}, where F is the intensity value at pixel position A. The bottom left panel shows wavelet result which displays temporal evolution of different periodicities. Power is plotted in inverted colours therefore, black indicates the region of strongest power. Cross hatched region is called cone of influence. This region can suffer from edge effects therefore, periods observed in this region are not reliable. Bottom right panel is the global wavelet plot which is the time average of the wavelet plot. The horizontal dashed line is the cutoff above which edge effects come into play and dotted line marks 95 \% significance level for a white noise process.  We find that significant power peaks at $\sim$52 sec followed by a second highest peak $\sim$13 s. The second peak is not significant and it could be due to small variations in light curve which are within the error limits.  Similar analysis is repeated for pixel position B where maximum significant power is $\sim$48 s and second peak $\sim$22 s (see Fig.~3 (b)). This analysis reveals that there is an indication of periodicities of around 15-30 s but they are not significant. Such periodicities could be the manifestation of small variation of intensities which are within error limits.\\
Fig.~3 (c) shows the wavelet map of AIA 193 \AA~ at pixel location marked as C (see Fig.~2 bottom right panel). We choose longer time series of ten minutes starting from 18:50:00 UT. Fig.~3 (c) top panel shows the AIA light curve with error estimates. Errors are calculated as $\sqrt{0.06F + 2.3}$ \citep{yuan2012}, where F is the intensity value at a given pixel position. Intensity values marked in red are co-temporal with Hi-C observations. We find that significant power peaks  $\sim$ 140 s. However, shorter periodicities are also present as small peaks at $\sim$ 59 s and 30 s. To investigate the presence of shorter periodicities below 60 s, we perform wavelet analysis of pixel location at C for the duration co-temporal with Hi-C observations as shown in Fig.~3 (d). We note that periods $\sim$ 60 s do exist but they are below significance level.\\It is worthwhile to note that power map reveals the finer structures in moss and braided magnetic region where power is concentrated.  In Fig~2 (bottom left panel) it is evident that the significant power (pixels marked in green) within 30-60 s are lying over active moss and braided magnetic field region in Hi-C FOV while significant power at these periods is absent in AIA 193 \AA~ (Fig.~2 (bottom middle panel)). However significant power at longer periods (120-180 s) are lying over active moss and braided magnetic field region in AIA FOV (Fig.~2 (bottom right panel)). It suggests that shorter periodicities (30-60 s) are present in smaller spatial scales which can not be resolved by AIA. Therefore, such short periodicites are almost absent in AIA 193 \AA~ FOV.
\subsection {Quasi periodic flows and Transverse Oscillation}
To study the flow within the braided magnetic region, we place three artificial curved slices along some specific threads as marked in Fig. 1~(e). Corresponding to each of three slices  a two dimensional time-distance diagram (x-t) is created (see Fig.~4), where x axis represent the time in seconds and y axis represent the distance along slit in Mm. Thick black region in x-t map (Fig.~4 top panel) represents the data gap. Ridges where structures can be resolved, gaussian curve along the column is fitted and the mean values with one sigma error bars is estimated. The ridge is then fitted with straight line. For extended or faint ridges the gaussian curve fitting is not feasible thus we fitted them visually with straight line. Fig.~4 bottom panel shows running difference x-t maps corresponding to maps in Fig.~4 top panel.\\
Several ridges are seen in x-t maps with different slopes at different times and with different lifetime as shown in Fig. 3. Slopes of the ridges gives an estimate of velocities of plasma outflow. The plasma outflows are found to be quasi periodic with large range of velocities from 13 km s$^{-1}$ - 185 km s$^{-1}$ (see Fig.~4) within our ROI which could be closely associated with reconnection sites .\\
Due to high spatial resolution Hi-C also helps us to probe transverse oscillations associated with fine threads. For that we place  three transverse slices across one highly braided structure as marked in Fig. 1~(f). For transverse slices, method of analysis is same as we adopted for curved slices. 
The x-t map (Fig.~5) is fitted with harmonic curve represented by the formula
\begin {equation}
y=a+bsin(\omega t + \phi) ,
\end {equation}
where $a$ is constant, $b$ is amplitude, $\omega$ is the frequency and $\phi$ is the phase.
 The amplitude of oscillation is found to be 110$\pm$109 km (111$\pm$102 km and 67$\pm$105 km) and period of oscillation is found to be 73$\pm$33 s (63$\pm$17 s and 53$\pm$22 s) for first (second and third slice). The velocity amplitude is found to be 9.4$\pm$13 km s$^{-1}$ (11$\pm13$ km s$^{-1}$ and $8\pm15$ km s$^{-1}$ ). We note  significant brightening along the loop, quasi periodic flows and transverse oscillations occur almost at same time for curved slice 1. The large error bars in velocity is because of large errors in the amplitude measurement. Errors in the amplitude estimation is large because one sigma error bars in the x-t map columns are larger than the observed amplitude of structure. Since we have aligned the data with sub-pixel accuracy so the amplitude observed is due to the displacement of structure along the slit and not due to shifts and jitter \citep{morton13,morton14}. Therefore, We find the evidence of short period, large velocity amplitude oscillations associated with braided magnetic structure.

\section{Conclusion}
In this study we investigate the high frequency dynamics of braided magnetic region as seen from Hi-C. Our filtered images reveal finer structures in contrast to the original images. These filtered images allows us to focus on the braided region and identify the threads within them. We produce power maps for periods in the 30-60 s interval and find that such periodicities are present in small spatial scales and can be detected with high temporal and spatial resolution only. Furthermore the power maps reveal the finer structuring within the moss region. We find that significant power is present in active moss and braided region as revealed by Hi-C. One can conjecture that high frequency power if they are due to waves are very much localized in the magnetic structures and are probably using them as wave guides for propagation. They seem to be restricted within these structures only. \citet{testa13} reported that the intensity variation in moss region is due to coronal nanoflares and estimated its timescale to be $\sim$ 15-20 s. In this study wavelet analysis reveal that shorter periodicities ($\sim$ 15-30 s) do exist but they are within the 95 \% significance level for a white noise process. Therefore, we can not distinguish white noise with periodic signal variation. We find periodic intensity variation with time period of 30-60 s above significance level which could be due to recurrent plasma outflows associated with repeated magnetic reconnection in these regions or due to  magnetoacoustic waves propagating along these fine structures which can generate periodic variation in intensity. It is worthwhile to note that we can not distinguish between flows or waves without spectroscopic observations. Therefore, in this study by flow we mean apparent motion in plane of sky.\\
Braided magnetic region as shown in Fig.~1(f) is known to manifest sufficient amount of magnetic free energy to heat the corona and to explain the observed brightening \citep{thalmann2014}. \citet{tiwari2014} reported subflares occurring in this region by analysing intensity in AIA images. They showed that such subflares are the consequences of magnetic reconnection. \citet{Cirtain13} reported outflow velocities to be 100-150 km s$^{-1}$ in this region driven by plasma pressure along the field as the plasma is heated locally because free energy released due to magnetic field dissipation due to reconnection. We have focused our attention to a highly probable reconnection site (see \citet{tiwari2014}). We find that the reconnections occurring in this region can drive plasma to a wide range of velocities from 13 km s$^{-1}$ - 185 km s$^{-1}$ projected in plane of sky depending on the strength of reconnection. Outflow velocities for most of them are lower than sound speed (100 km s$^{-1}$ for 1 MK plasma) while some of them are greater than sound speed. \\
\citet{morton13,morton14} reported the period of transverse oscillations in moss region peaking around 50 s, displacement around 40 km and velocity amplitude peaking around 3 km s$^{-1}$. In this study we focussed on short period and large amplitude oscillation and we find that short period and large velocity amplitude oscillations are associated with braided magnetic region where reconnection happens \citep{thalmann2014}. \citet{mc2011} reported the velocity amplitude of 5$\pm$5 km s$^{-1}$ using Monte-Carlo simulation in active region of the Sun at coronal heights. Such low frequency waves are not energetic enough to heat active corona. They have conjectured the presence of shorter period waves in active region which can not be resolved by SDO/AIA. Our analysis confirms the presence of large amplitude ($\sim$ 8--11 km s$^{-1}$) high frequency ($\sim$53--73 s) transverse waves in braided magnetic region. The amplitude of these oscillations are $\sim$67--110 km which is far below the spatial resolution limit of AIA. Therefore such oscillations can not be resolved by AIA. Such waves could be responsible for heating active corona. Since significant brightening and plasma outflows observed in braided magnetic region and transverse oscillation happens at same time it is not yet clear if reconnection is driving the transverse waves or if transverse waves cause magnetic field lines to reconnect. Further studies are required to confirm this. Our results confirms the necessity of high temporal and spatial resolution combining imaging with spectroscopy for future missions. 





We acknowledge the High resolution Coronal imager instrument team for making the flight data publicly available. MSFC/NASA led the mission and partners include the Smithsonian Astrophysical Observatory in Cambridge, Mass; Lockheed Martin Solar Astrophysical Laboratory in Palo Alto, Calif; the University of Central Lancashire in Lancashire, UK; and the Lebedev Physical Institute of the Russian Academy of Sciences in Moscow.

\clearpage



\clearpage









\clearpage



\clearpage



\clearpage
\begin{figure}
\label{sun}
\figurenum{1}
\begin{center}
\includegraphics[scale=0.5]{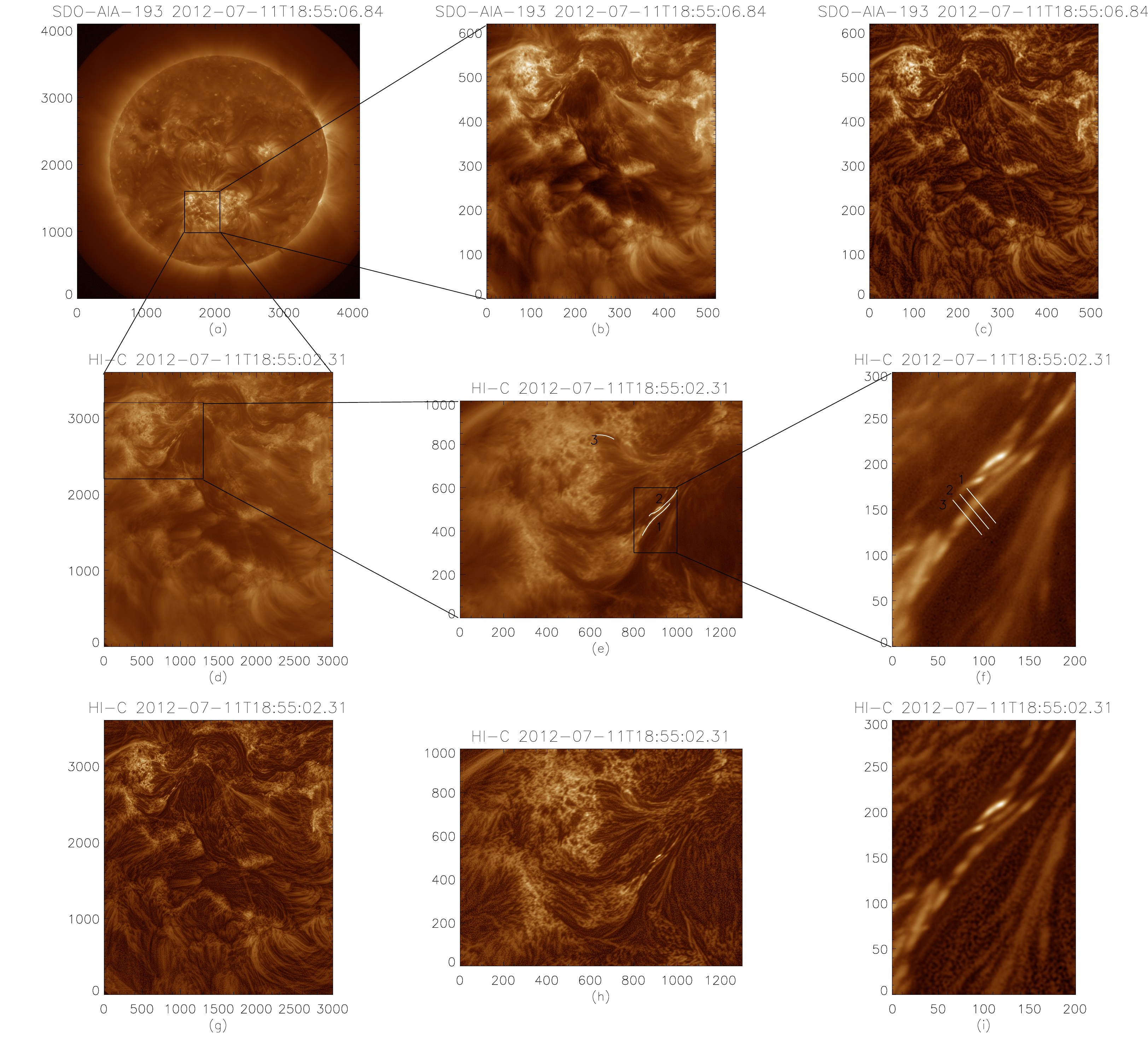}

\caption{{\it Top panel}: (a):AIA 193 \AA~ full disk image. Black rectangle represents the full Hi-C FOV. (b): AIA 193 \AA~ having same FOV as Hi-C. (c):Reconstructed image, using multi gaussian normalized filter, with the same FOV as shown in (b). {\it Middle Panel}:(d): Full FOV of  HI-C image. Region under consideration is indicated with black rectangle. (e): Three curved slits positions, indicated by white curves,  are shown which are used for time distance maps. (f): Three slits are placed perpendicular to the threads as indicated with white lines for the detection of transverse motions.
{\it Bottom panel}: Reconstructed images, using multi gaussian normalized filter, with the same FOV as shown in middle panel. 
}
\end{center}
\end{figure}

\begin{figure}
\label{power}
\figurenum{2}
\includegraphics[scale=0.5,angle=90]{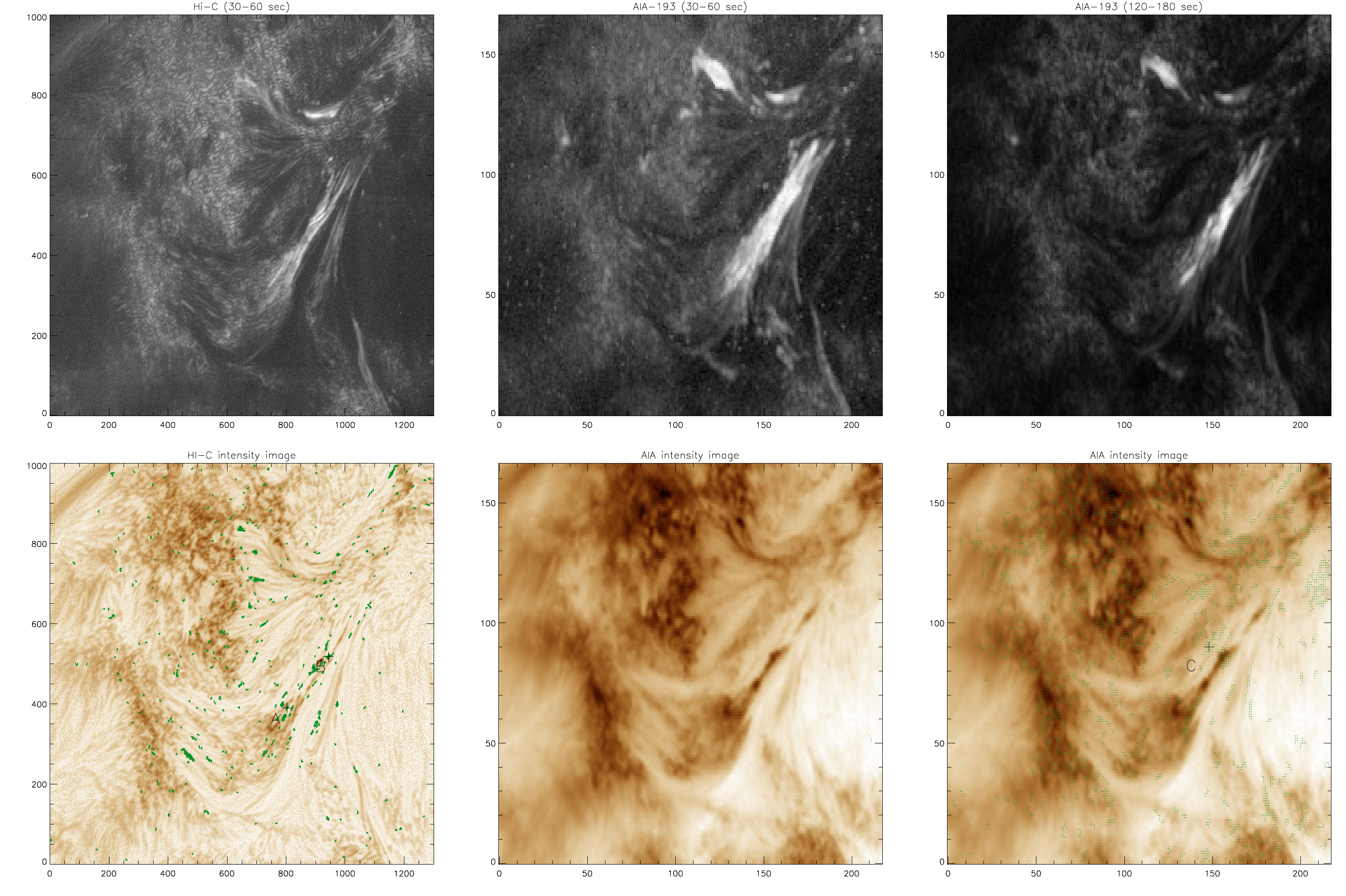}

\caption{{\it Top panel}: left: Power map of Hi-C region of interest plotted in logarithmic scale for 30-60 s. Middle: Power map of AIA 193 \AA~ region of interest plotted in logarithmic scale for 30-60 s periods. Right: Power map of AIA 193 \AA~ region of interest plotted in logarithmic scale for 120-180 s. 
{\it Bottom panel}:  left: Filtered inverted intensity image of Hi-C overplotted with pixels positions having significant power greater than 95$\%$ confidence level in green.  Middle and Right: inverted intensity image of AIA 193 \AA~ overplotted with pixels positions with significant power greater than 95$\%$ confidence level in green.
}
\end{figure}

\begin{figure}
\label{wavelet}
\figurenum{3}

\subfigure[]{\includegraphics[bb=0 0 350 700,clip,scale=0.35,angle=90]{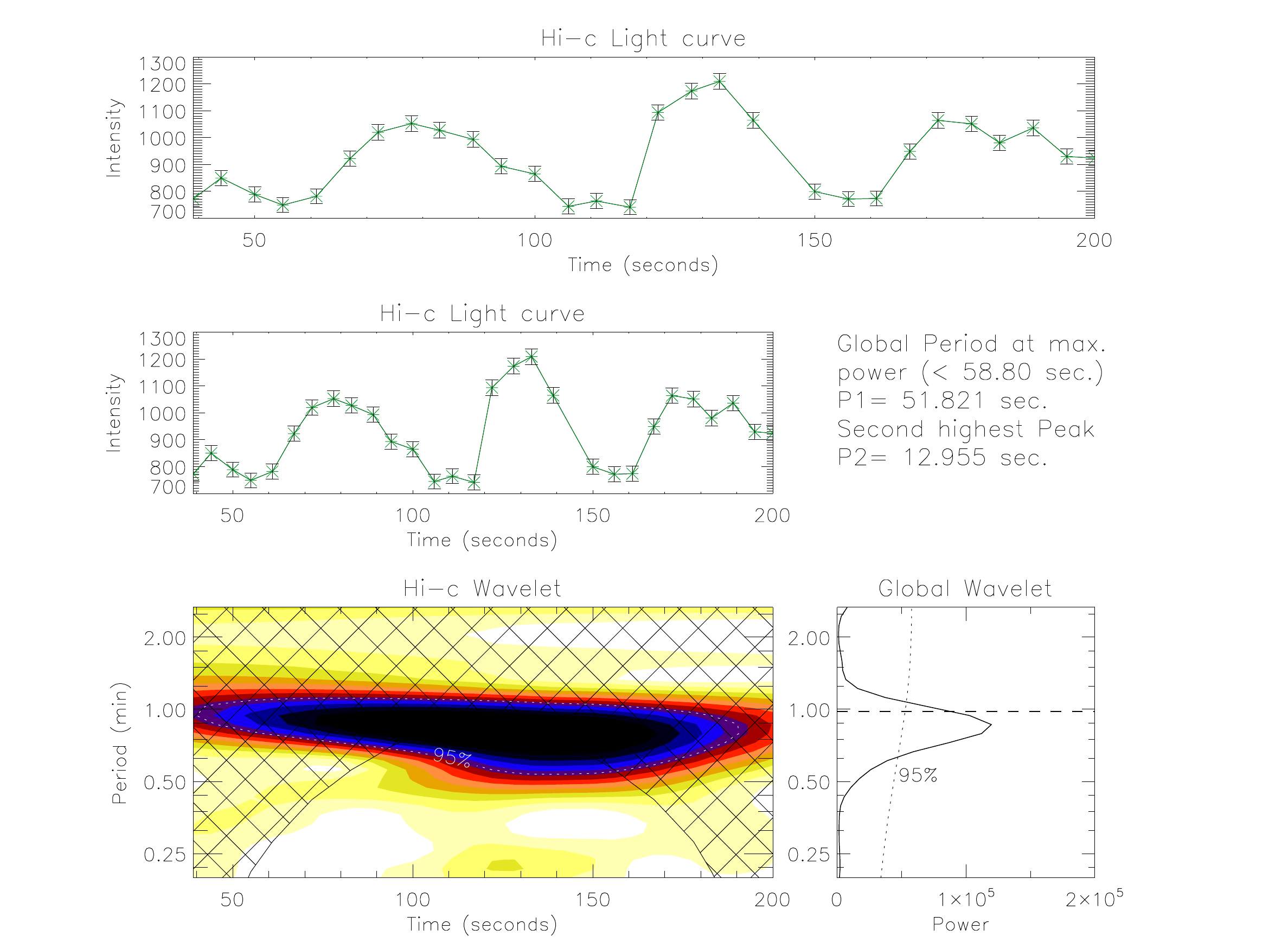}}
\subfigure[]{\includegraphics[bb=0 0 350 700,clip,scale=0.35,angle=90]{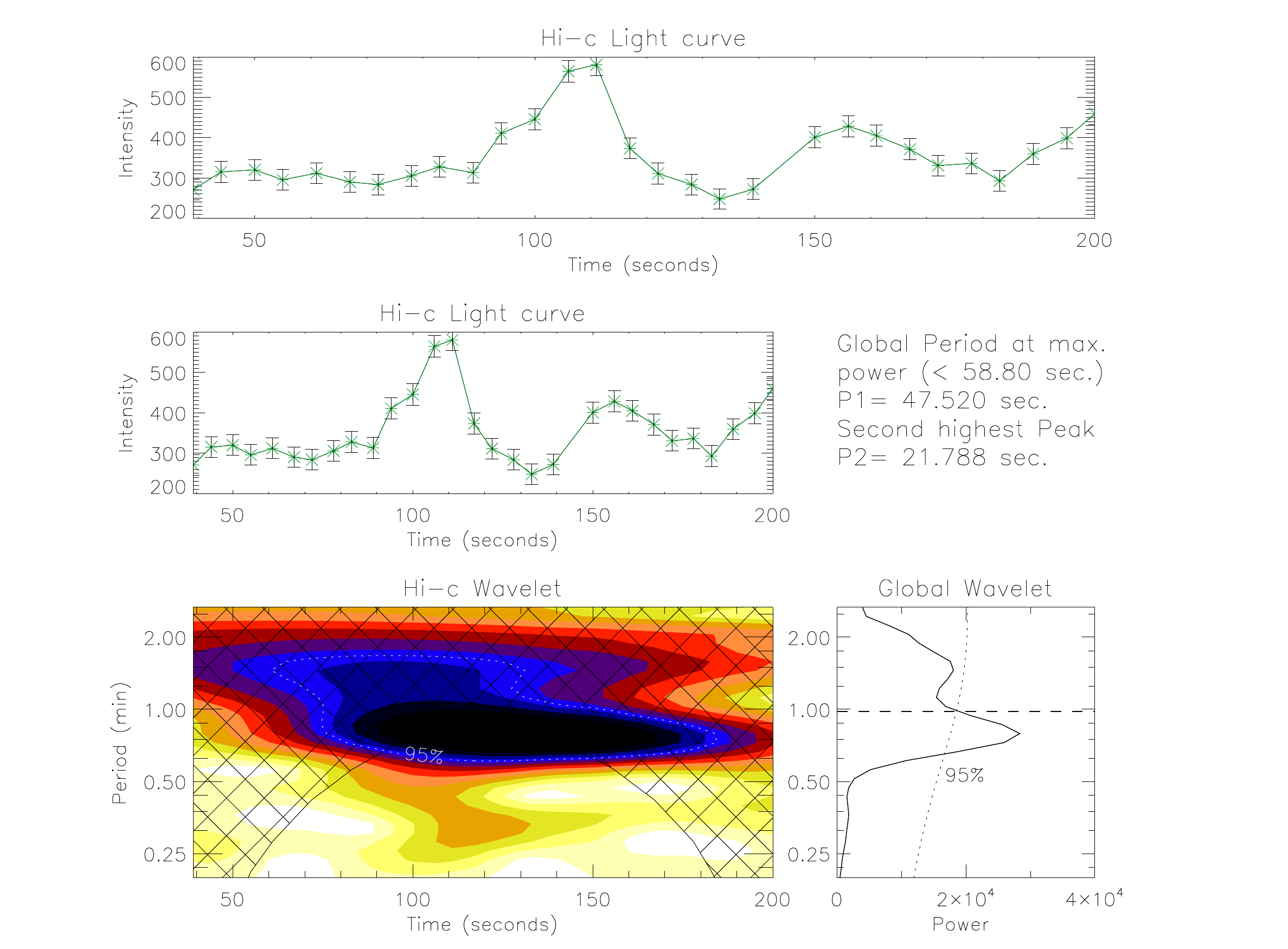}}
\subfigure[]{\includegraphics[bb=0 0 350 700,clip,scale=0.35,angle=90]{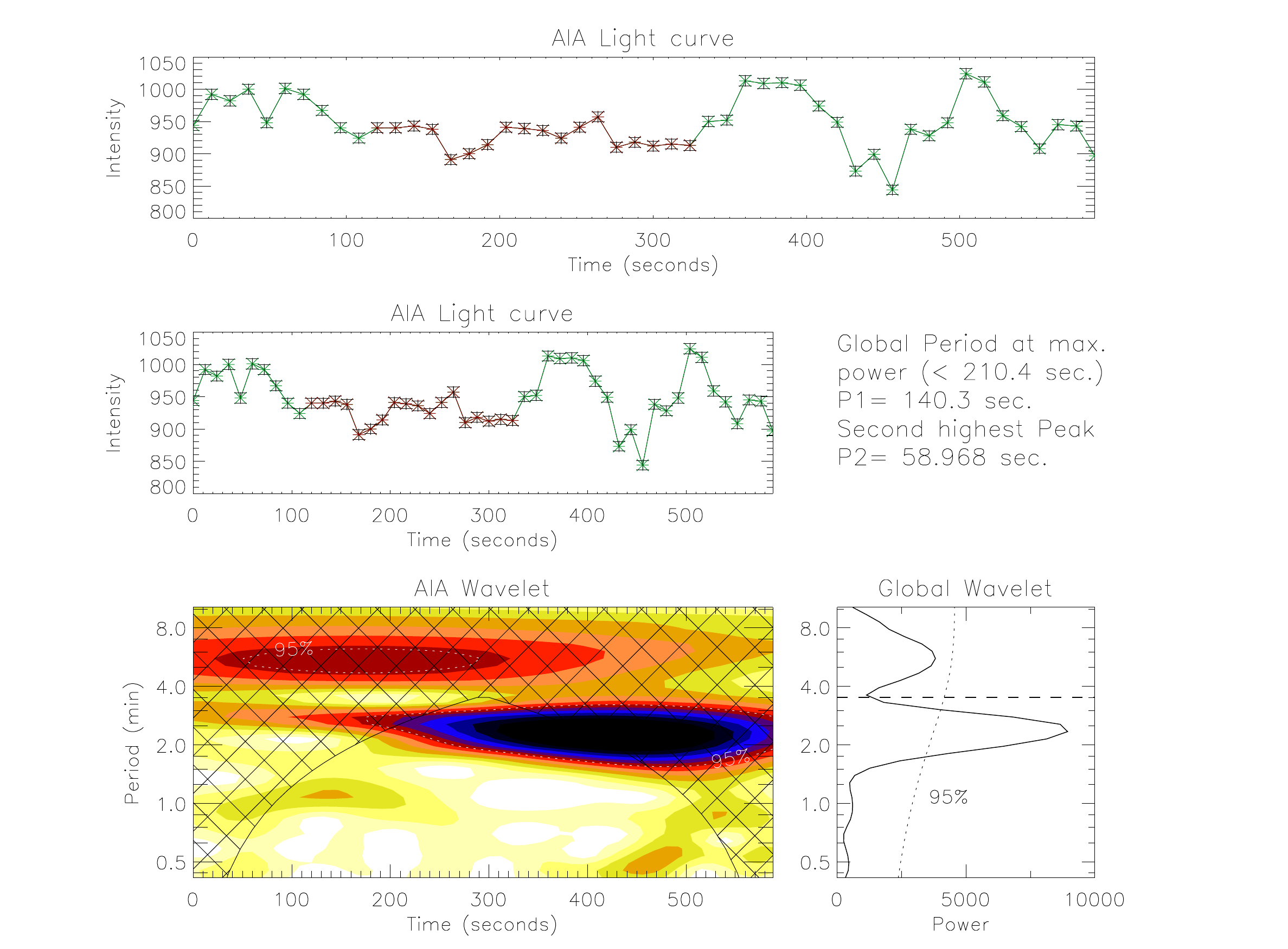}}
\subfigure[]{\includegraphics[bb=0 0 350 700,clip,scale=0.35,angle=90]{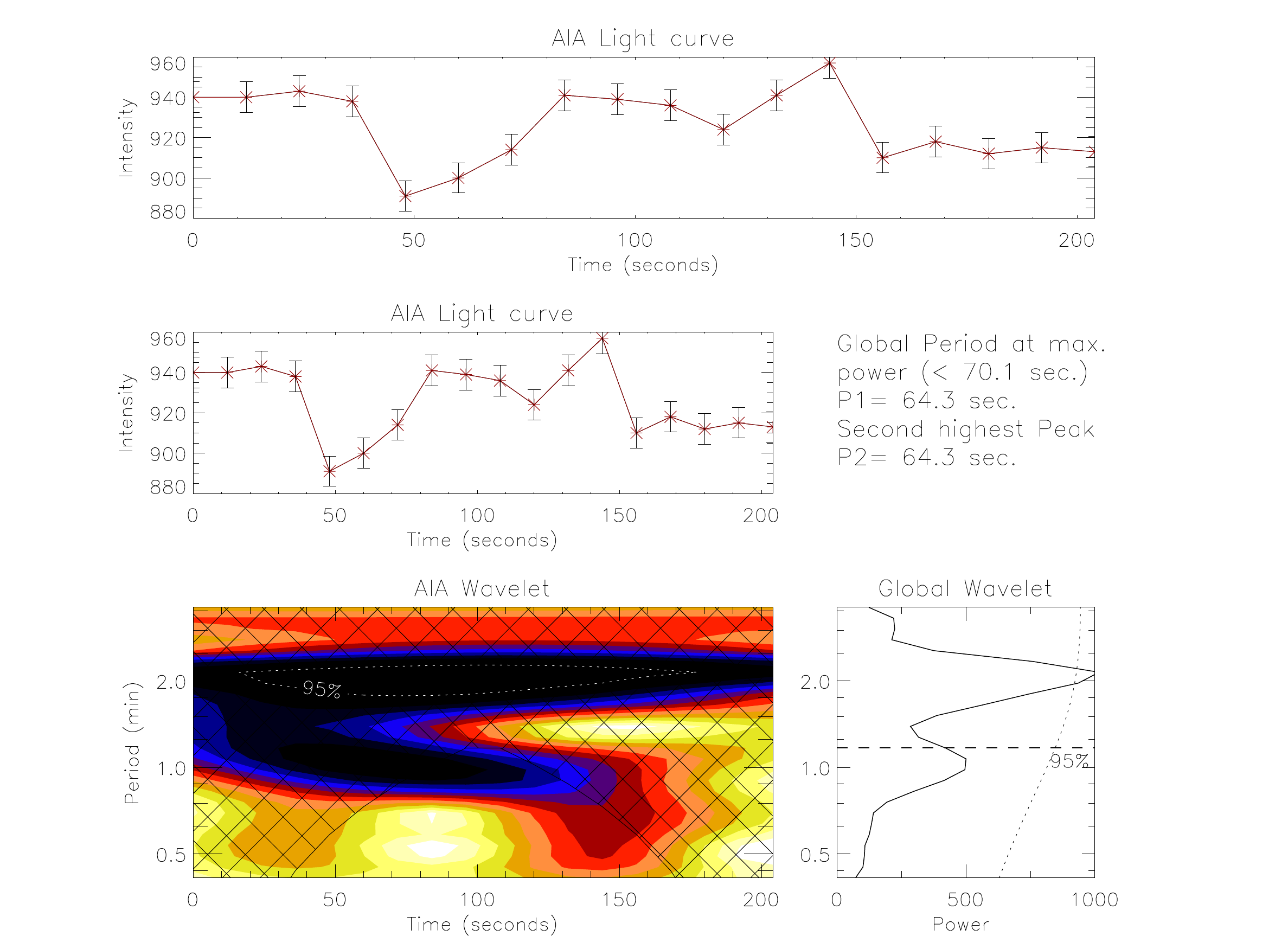}}
\caption{(a): Wavelet map of Hi-C intensity at pixel position `A' marked in Fig.~2 bottom left panel. (b): Wavelet map of Hi-C intensity at pixel position `B' marked in Fig.~2 bottom left panel. (c): Wavelet map of AIA intensity at pixel position `C' marked in Fig.~2 bottom Right panel. Data points marked in red are co-temporal with Hi-C duration. (d): Wavelet map of AIA intensity at pixel position `C' marked in Fig.~2 bottom Right panel co-temporal with Hi-C duration. 
}
\end{figure}

\begin{figure}
\label{long}
\begin{center}
\figurenum{3}
\plotone{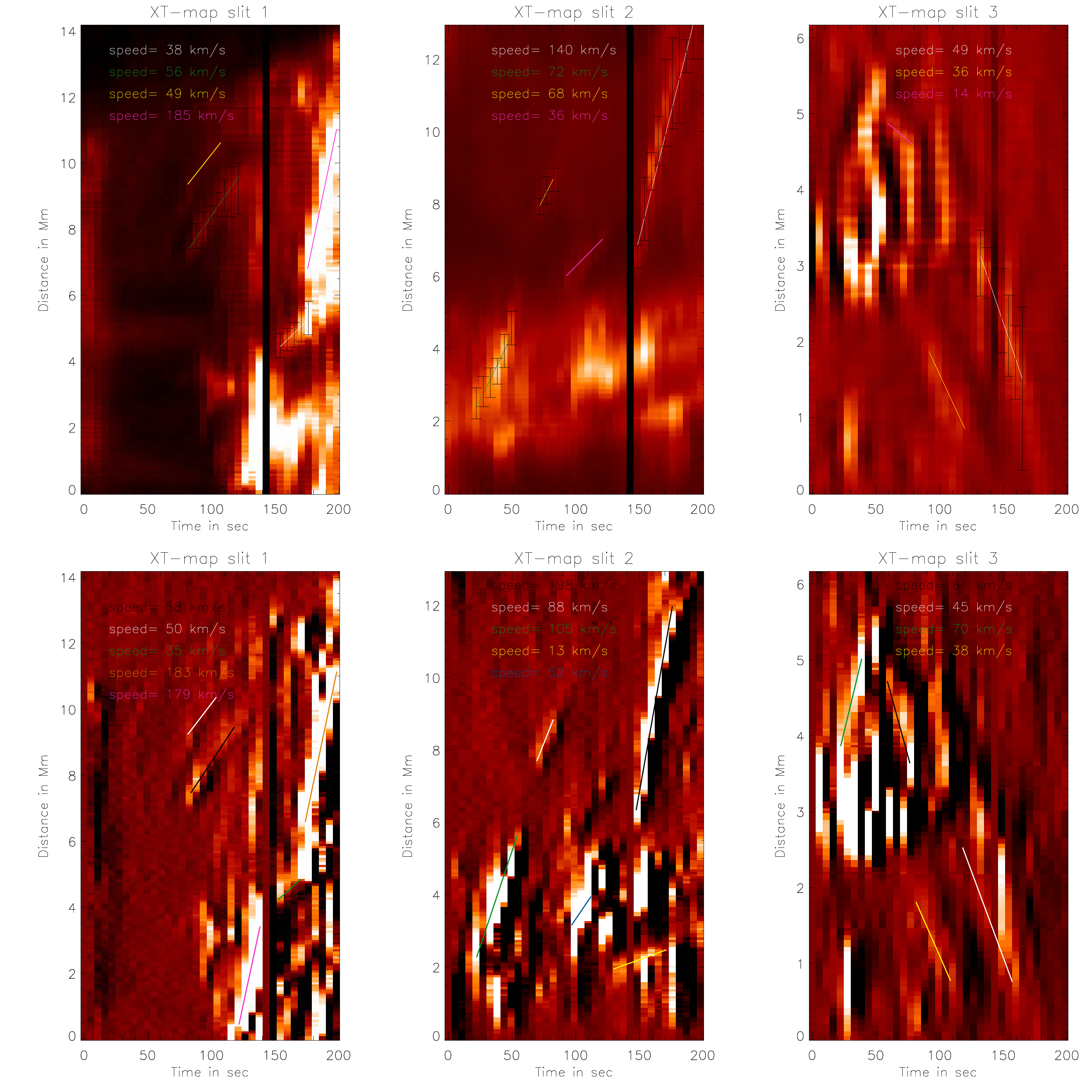}

\caption{{\it Top panel}: Time - distance (x-t) maps corresponding to the three curved slices  marked as 1, 2 and 3 in Fig.~1(e) overplotted with best fit straight lines on the observed ridges. The slopes of the fitted lines provide an estimate of the speeds are also printed. Black vertical strip represent the data gap of one frame. {\it Bottom panel}: Time - distance (x-t) maps of running difference images corresponding to x-t maps in top panel.} 
\end{center}
\end{figure}

\begin{figure}
\label{trans}
\figurenum{4}
\plotone{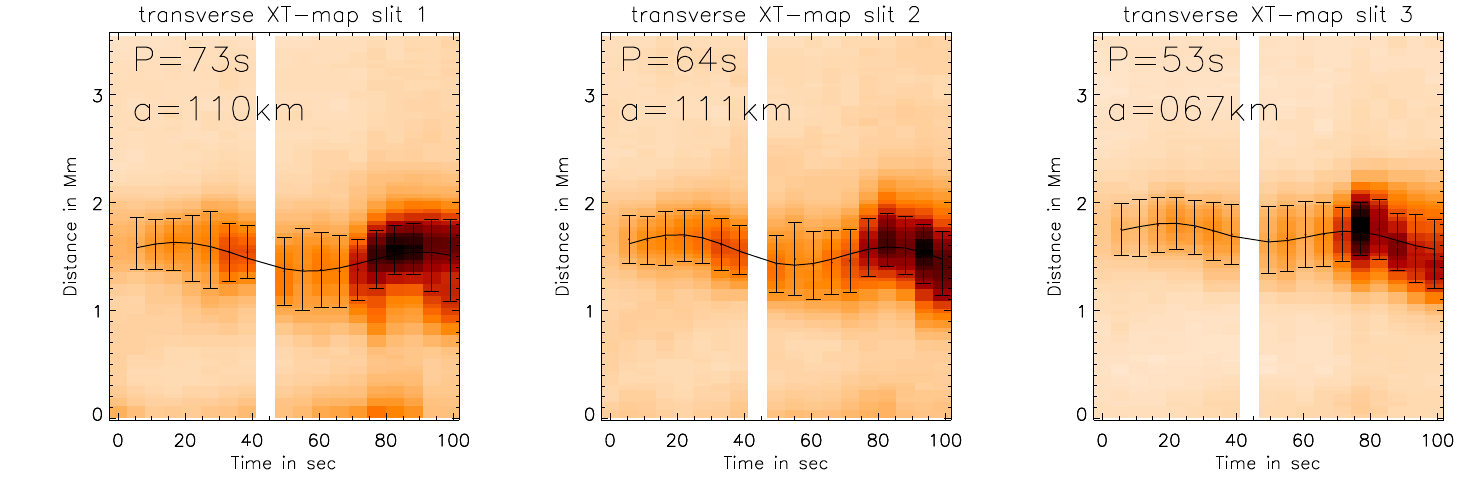}

\caption{Inverted intensity x-t maps corresponding to the transverse slices marked as 1, 2 and 3 in Fig.~1(f) overplotted with best fit sine curves. P and a are time period and amplitude of oscillations. White vertical strip represent the data gap of one frame.} 
\end{figure}




\end{document}